\documentclass[11pt, a4paper]{revtex4}
\usepackage{graphicx}
\usepackage{dcolumn}
\usepackage{latexsym}
\usepackage{amsmath, amsthm, amssymb}
\usepackage{epsfig}
\usepackage{bm}
\usepackage[dvipsnames]{xcolor}

\begin{document}
\title{Dynamics of a particle moving in one dimensional Lorentz lattice gas}

\author{Sameer Kumar}
\email[]{sameerk.rs.phy16@itbhu.ac.in}
\affiliation{Indian Institute of Technology (BHU), Varanasi, U.P. India - 221005}

\author{Shradha Mishra}
\email[]{smishra.phy@itbhu.ac.in}
\affiliation{Indian Institute of Technology (BHU), Varanasi, U.P. India - 221005}

\begin{abstract}
We study the  dynamics of a particle moving in one-dimensional Lorentz lattice-gas where particle performs mainly three different kinds of motion {\it viz} ballistic motion, 
diffusion and confinement. There are two different types of scatterers, {\it viz} reflector and transmitters, randomly placed in the lattice.  Reflectors are such that they reverse the particle's velocity direction and transmitters let it pass through. Scatterers also change their character with flipping probability $1-\alpha$, once the particle interacts with a scatterer. Hence the system is defined by two sets of parameters, $r$, which is the initial density of reflector/transmitter and $\alpha$.  For $\alpha=0$ and $\alpha=1$ dynamics of the particle is purely deterministic else it is probabilistic. In the pure deterministic case dynamics of the particle is either propagation in one direction or confined between two near-by reflectors present.  For the probabilistic case  $\alpha \ne 1$ and $\ne 0$, although the dynamics of particle shows anomalous diffusion where dynamics is faster, slower and comparable to normal diffusion on the variation of system parameters $(\alpha, r)$, but the asymptotic behaviour of the particle is normal diffusion. We plot the phase diagram for the asymptotic behaviour, in the plane of $\alpha$ and $r$.
\end{abstract}

\maketitle
\section{Introduction}

Most natural micro-swimmers usually move in a complex environment and encounter soft and solid walls, obstacles, i.e. a {\it heterogeneous} environment \cite{goldingprl2006, naturemater2013, parry}, which can be realised by regular or irregular patterns of obstacles, which control their motion depending on the background environment.  Their dynamics can vary
from confined trajectories, subdiffusion, diffusion, super diffusion to propagation \cite{suryanjop2016, revarticle}. How does the nature of surroundings affect the dynamics are the questions addressed in many studies \cite{stark, subdiffusion}. One of the ways to model motion of a particle in such a complex environment is through Lorentz lattice gas \cite{205llg, binder1987, binder1988}.
In the past, there have been a number of ways in which a Lorentz lattice gas (LLG) has been used to model different physical phenomena \cite{langton, gale, troubetskoy, beijeren1982}. Most of these studies are for two and higher dimensions \cite{shradhajstatphys, meng1994, benweb, kong1991}. But study  in one-dimension is also interesting which can help us to understand the dynamics of many one dimensional systems like: ant moving on a trail \cite{ant}, motion of motors on filament \cite{motor}, transport of proteins along the channel \cite{protein} etc.. Also, a one dimensional model has less number of control parameters and hence give more insight to the system. 

In a Lorentz \cite{langton} lattice gas (LLG) a single particle moves along the bonds of the lattice. When it arrives at a lattice site, the particle encounters a scatterer, which scatters the particle according to some fixed rule. In addition to the particle, each scatterer can also have some different orientations, or more generally states, that may also change over time as it interacts with the particle, etc. Hence a Lorentz lattice gas is defined by (i) underlying lattice (ii) the initial density of scatterer, which is called as the LLG’s initial configuration. One of the main question we ask here is how do the dynamics of particle change as we change the density and property of the scatterers.

In our present study, we introduce a one-dimensional lattice of unit lattice spacing $a=1$ on which two types of scatterers, ``reflectors" and ``transmitters", are present; and both are randomly distributed. A reflector reverses the direction of the particle's velocity, and a transmitter lets the particle pass through. The density of scatterers is controlled by a number $r$ which is defined as $r=\frac{C_{R}}{C_{L}+C_{R}}$; where $C_{R}$ and $C_{L}$ are the initial concentration of reflector and transmitters respectively. Reflectors ($R$) and transmitters ($T$) also  flips i.e. $R \leftrightarrow\ T$ and $T \leftrightarrow\ R$ with a probability $1-\alpha $, once the particle pass through. For $\alpha = 0$, there is always flipping, and for $\alpha\ = 1$  there will be no flipping (i.e. fixed); hence, dynamics is deterministic. And for $0<\alpha\ <1$ the flipping is probabilistic.  Any finite $\alpha \neq 1, 0$ makes model complex, and we conclude the results through numerical study. But for $\alpha = 0$ and $\alpha = 1$, we propose some analytical argument, which, matches with previous results \cite{beijeren1982}.

Now we will briefly discuss our main results. We study the model for range of $\alpha \in [0,1]$ and density $r\in [0,1]$. For $\alpha=0$ (pure flipping) dynamics is always ballistic, and the direction of the particle velocity depends on the configuration of scatterers in the system. When $r = 0.0$ and $r=1.0$ (only one kind of scatterers), the particle moves in the direction of its initial velocity. When there are both kinds of scatterers in the system (i.e. when $r \neq 0.0$ and $r \neq 1.0$) and its first encounter a reflector,\ followed by a transmitter then the direction of the motion of particle will be opposite to that of its initial velocity. In another case, if the particle encounters a transmitter at the start of the motion, on average, it will continue to move in the same direction that of its initial velocity. Also for the pure flipping case speed of the particle decreases with an increase in $r$  as $v=\frac{1/r}{1/r +2}$. For a purely fixed case, when $\alpha=1$, dynamics of the particle always confined between two nearest reflectors present in the system and span of confinement varies with $r$. In this case the spread of the confined region i.e. radius of gyration $R_g (=\sqrt{\Delta(t)})$ linearly varies with $1/r$. For the case when $\alpha \ne 0$, $\ne 1$: the dynamics of the particle is probabilistic. We plot the full phase diagram for the asymptotic behaviour in the plane of ($\alpha$-$r$), which is in a good agreement with some of the phases observed for particle motion by Kong {\it{et al}.} \cite{kong1991}. Dynamics is characterised by mean square displacement (MSD) exponent $\beta$, such that at late time MSD, $\Delta(t) \propto t^{\beta}$. For fixed $r$, decreasing  $\alpha$ from $0$ to $1.$, i.e. going from pure flipping to fixed, the exponent $\beta$ decreases from $2$ (ballistic) to $0$ (confinement). In general, approach to the asymptotic behaviour happens through mainly two states over time where the motion continues to progress in the same fashion as the early time behaviour like in the case of ballistic motion, and in other cases, the dynamics is initially faster and then slows down to show normal diffusion. But when $\alpha$ is close to $0$, i.e. $0+\delta$ where $\delta \simeq 0.001$ then the dynamics approaches its asymptotic behaviour mediated by three regimes which are further explained in the results section \ref{results}. In our model full range of dynamics can be seen by varying the two parameters $\alpha$ and $r$. Changing the two parameters system shows a transition from one type of motion to another. 

The rest of the article is divided in the following manner. In the next section \ref{model}, we first describe our model. In section \ref{definition}, we defined the three
types of motion and then in section \ref{results},  we discuss our results in detail and finally conclude in section \ref{discussion}.
\\
\begin{figure}
\centering
\includegraphics[width=8cm,height=6cm]{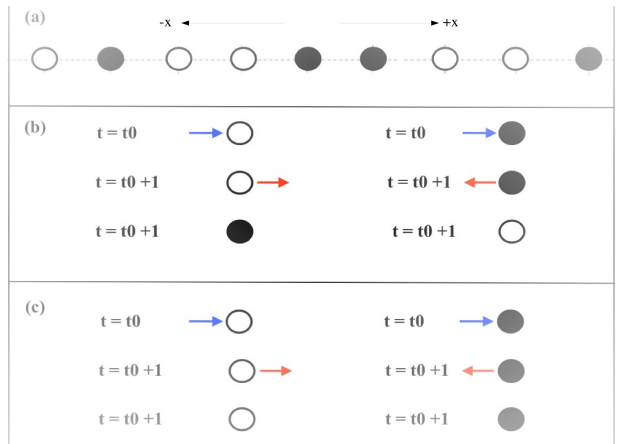}
\caption{A cartoon of  one dimensional Lorentz lattice Gas : filled circles:- reflector, empty circles:- transmitters. Arrow shows the direction of particle's velocity. (a) Part of a typical initial configuration. Interaction of particle with a scatterer and state of scatterer  one-step after interaction when (b)  $\alpha=0$ i.e. pure flipping and (c) $\alpha=1$ i.e. pure fixed.} 
\label{fig: 1}
\end{figure}
\\
\section{Model and numerical details}
\label{model}

In our model, a single particle moves along the bonds of the lattice of unit lattice spacing $a=1$ in the unit time step, i.e. $(\Delta t = 1)$. In the lattice, the two types of scatterers: ``reflectors'' and ``transmitters'' are randomly distributed. Reflectors reverse the direction of the particle's velocity, and transmitters let the particle move in the same direction. The reflectors ($R$) and transmitters ($T$) also flips (after the particle pass through) with probability $1-\alpha$ ($\alpha \in [0,1]$). If $\alpha\ = 0$ there will be always flipping and if $\alpha\ = 1$ there will be no flipping, and if $0 < \alpha\ < 1$ then flipping will be probabilistic. We vary  the initial  density  of reflectors $R$ and transmitters $T$ according to a  number $r$, such that if $r = 0$, initially all the scatterers are transmitters;  if $r = 1$, all are reflectors, and if $r = 0.5$, scatterers are in equal ratio. A cartoon picture of part of the model shown in fig. \ref{fig: 1}.

We start with a random initial distribution of $R$ and $T$  in the lattice. One of the typical initial configurations of $R/L$ shown in fig. \ref{fig: 1} (a). A particle starts to move along a randomly chosen direction, forward ($+x$ direction) or backward ($-x$ direction), from the centre of the lattice and move along the bonds of the lattice. The direction of the particle's velocity changes according to the presence of $R$ or $T$ at each lattice site. For example, if the particle encounters a $R/T$, then its velocity direction will switch back/remain the same (reflected/transmitted). At the same time, $R/T$ will change to $T/R$ with probability $1-\alpha$. Hence initial configuration of $R/T$ is going to change with the dynamics of the particle. Dynamics of the particle explored for the various choices of the initial concentration of $R/T$, i.e. $r$ and flipping probability $1-\alpha$. The initial configuration generated for a large enough but fixed lattice size such that the particle never reaches the boundary. Hence boundary plays no role. Properties of the system is characterised by calculating (a) Mean square displacement, MSD of particle position defined as $\Delta_{\alpha, r} \left( t \right)\; =\; <\left[ x\left( t \right)\; -\; x\left( 0 \right) \right]^{2}>$, where $<..>$ denotes the average over many initial realisation of $R/T$ and for a given choice of $r$ and $\alpha$. (b) Number of different visited sites $N(t)$, (c) density of scatterers on the visited sites $r_{visited}(t)$ and (d) Probability distribution of particle position $P(x,t) $. Also at a long time when the motion is diffusion $N(t)$ should satisfy the equation \ref{eq:1},

\begin{equation}
\centerline{$\frac{d}{dt}N(t)=\frac{c}{N(t)}$}
\label{eq:1}
\end{equation}

where the value of $c$ depends on the system parameters $\alpha$ and $r$.

\begin{figure}
\centering
 \includegraphics[width=12cm,height=8cm]{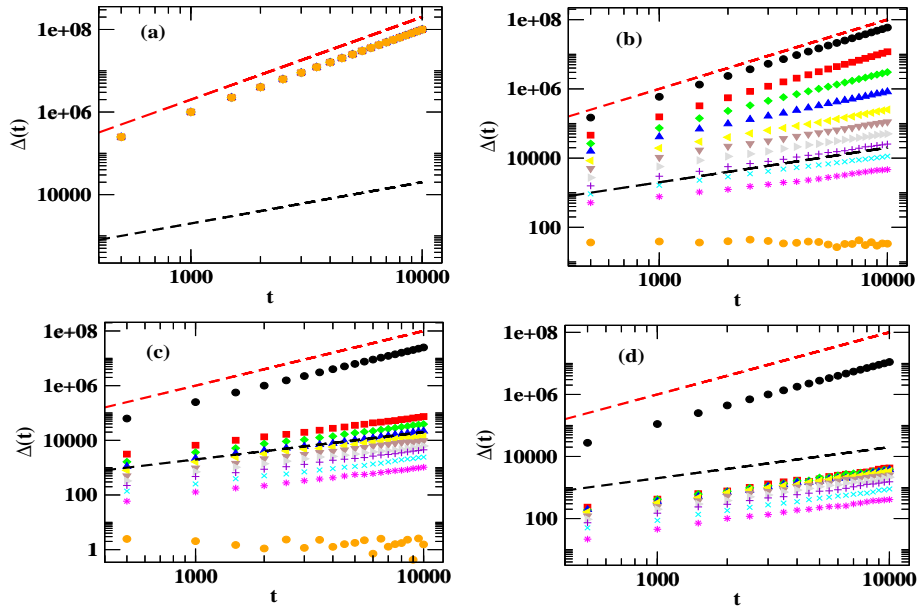}
 \caption{(Color online) MSD $(\Delta(t)$ {\it vs.} time $(t)$  (upto $10^4$ before the asymptotic behaviour observed) Plot: $r=0.0$ (a), $r=0.15$ (b), $r=0.5$ (c) and $r=1.0$ (d) for $\alpha$ = 0.0 (circle), 0.1 (square), 0.2 (diamond), 0.3 (triangle up), 0.4 (triangle left), 0.5 (triangle down), 0.6 (triangle right), 0.7 (plus), 0.8 (x), 0.9 (star), 1.0 (open circle). Black and Red dotted lines have slope = 1 and 2 respectively. Data is averaged over 1000 ensembles.}
 \label{fig: 2}
\end{figure}
Before we discuss our results, we first define the three kinds of motion in the next section \ref{definition} and then we discuss the result for different values of $\alpha$ and $r$ in section \ref{results}.   
\section{Definitions: Ballistic, Anomalous diffusion and Confined motion}
\label{definition}
\begin{figure*}
\centering
\includegraphics[width=12cm,height=8cm]{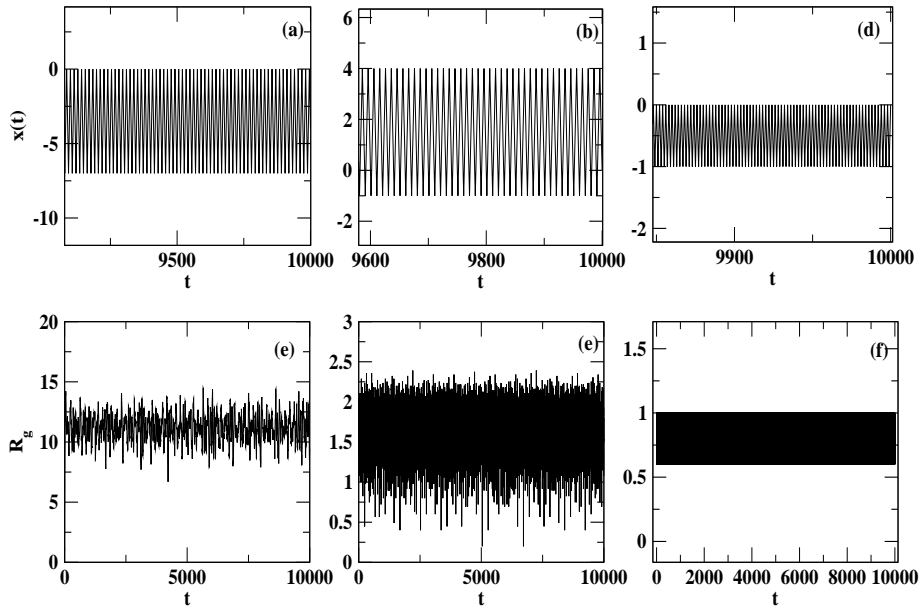} 
\caption{Three different trajectories when particle motion is confined (or periodic): (a), (b) and (c) shows the trajectory for system parameter set ($\alpha$, r)=(1.0,0.1), (1.0,0.5) and (1.0,0.9) respectively. Figure (d),(e) and (f) shows the RoG of the particle for system parameter set ($\alpha$, r)=(1.0,0.1), (1.0,0.5) and (1.0,0.9) respectively.  Data for $R_{g}=\sqrt{\Delta(t)}$ is averaged over 100 realisations.}
\label{fig: 3}
\end{figure*}

 We characterise the three different kinds of motion by calculating the MSD. In general, with time, MSD varies as $\Delta\left( t \right)_{\mbox{lim}_{t\rightarrow \infty}} \simeq t^{\beta}$, where we define, $$\beta = \lim_{t \rightarrow \infty} \frac{ln \Delta (t)}{ln(t)}$$ as the MSD exponent. All the measurements are done in the asymptotic state when $\beta$ approaches a constant value at the late time (i.e. remain constant for at least two decades). In general, we can define an initial transient state in the system, i.e. for all set of parameters $(\alpha, r)$, the system shows a transition from early time transient state to late time steady state. Near to the phase boundary shown in phase diagram \ref{fig: 8}, the transition from the early time transient state to the late time asymptotic state appears after a long time ($>10^7$). 
We calculate the MSD i.e $\Delta(t)$ vs. $t$ and extract the exponent $\beta$. MSD vs time plot shown in fig. \ref{fig: 2} for some choice of values of $r (=0.0, 0.15, 0.5, 1.0)$ and $\alpha \in [0,1]$. The dynamics of particle is ballistic if $\beta=2$, hence particle on average moves in one direction with certain speed $v(t)=\sqrt{\Delta(t})/t$. The maximum possible speed of the particle can be $ v(t)=\frac{a}{\Delta t}=1$ when it always move in one direction. In general, the particle can spend some of its time moving forward and backwards, but on average moving in one direction. In that situations speed $v(t)$ is less than $1$. A ballistic motion of the particle happens when the particle does not scatter frequently but moves smoothly, i.e. have negligible resistance in the system. Or we can say that the reflectors in the system favour the motion of the particle to be in one direction. We call the dynamics of the particle to be normal diffusion if the  MSD exponent, i.e. ${\beta \simeq 1}$. We call the dynamics of the particle is confined if the exponent $\beta$ approaches zero. Trajectories for such kind of motion shown in fig. \ref{fig: 3} for different set of $(\alpha,r)=(1.0,0.1),(1.0,0.5),(1.0,0.9)$.

\section{Results}
\label{results}
Now we discuss our results in detail: 

{\bf Case I: Pure transmitter $r=0$}:- Initially when all the obstacles are of transmitter type, then whatever is the value of flipping probability $1-\alpha$ particle will always move in a straight line in the direction of its initial velocity with its maximum speed $v_{0}=1$. It is trivial because in this situation, when all of the obstacles are transmitter particle will always pass through and never come back, and hence, the value of $\alpha$ will be irrelevant. Hence  $\beta =2$ and $v(t)=1.0$ 

{\bf Case 2: Pure fixed}:- When $\alpha=1$, or obstacles never change their character (i.e. fixed type) then the dynamics of the particle will always be confined between two nearest reflectors. Hence statistically when averaged over a large number of initial realisation for a given $r$, average distance between the two nearby reflectors is $1/r$. Mean square displacement of the particle trajectory saturates to some finite value after an initial transient state. The square root of the MSD at the late time: i.e. in the stationary state determines the spread of the particle trajectory or also called as the radius of gyration  (ROG) $R_g = \sqrt{\Delta(t)}$ for large $t$, fig. \ref{fig: 3}(d, e and f). This result suggests that in confined motion particle explore the different amount of space for different value of $r$ and $R_g \sim \frac{1}{r}$.

{\bf Case 3: Pure flipping}:-When $\alpha=0$, i.e. when properties of obstacles are purely flipping type, then dynamics of the particle is always ballistic in the direction of initially chosen velocity
direction. But this case is different from {\bf case 1}, ($r=0$), and speed of the particle will depend on the concentration of $r$ on the lattice. It is a trivial exercise to check on a piece of paper that for this case particle will move on average $1/r$ distance in $1/r+2$ time steps. Hence the speed of the particle can be estimated to be $v=\frac{1/r}{(1/r)+2}$. As  $r \rightarrow 0$, speed $v$ approaches $1$ ({\bf case 1}) and as $r \rightarrow 1.0$, speed $v=1/3$,  which matches with the results of Grosfils {\it{et al.}} \cite{bunimovich1999}, where  velocity of particle moving in propagation mode (i.e. ballistic motion) in a model where a particle moves in a one dimensional lattice occupied by two different kinds of scatterers, {\it viz} $spin \ up$ and $spin \ down$.

{\bf Case 4: The stationary state}:- When $r=0.5$ i.e both types of scatterers are in equal ratio then the motion is always diffusion for all values of $\alpha$ except when $\alpha=0 \ or 1.0$. In this case, the particle has an equal chance to go left and right, and this is the case similar to the one-dimensional random walk. Also, this is the case when the system remains in stationary state irrespective of the time particle have spent in the system. Therefore for other value of $r$, asymptotically, the system approaches a stationary state in which $r_{visited}=0.5$. To understand this, consider the situation that the particle sits on a new site, $i$, visited for the first time. The probability that it will return to this site is either unity or less, say $\rho$. In the latter case, the walker has to move ballistically in either direction, in $1$ out of $(1-\rho)^{-1}$ newly visited sites the process renews itself; it never returns to the last site it came from.  In the former case, the walker will keep returning forever (and the environment of a given site becomes more diffusive and less ballistic with every return). After a number of visits of order $\frac{1}{\alpha}$ the neighbourhood of site $i$ becomes equilibrated, implying $r_{visited} = 0.5$.\\

\begin{figure}
\centering
\includegraphics[width=15 cm,height=4.0 cm]{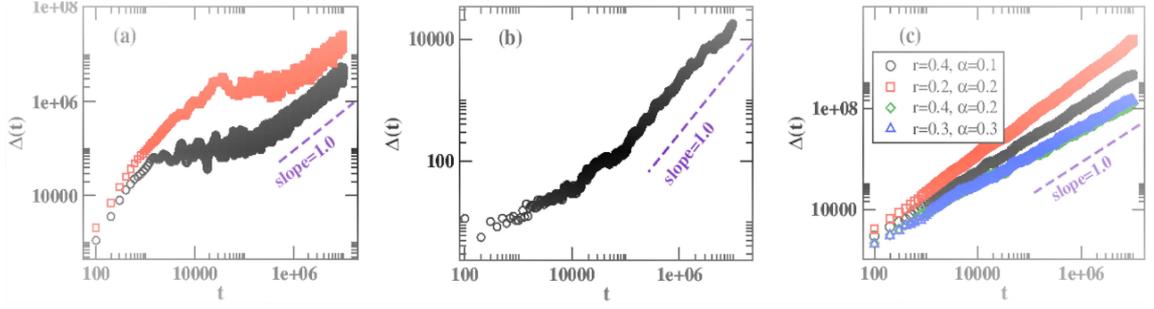}

\caption{MSD ($\Delta(\tau)$) vs. time ($\tau$) plot  when (a) $\alpha = 0.003$ and $r=0.9 \ and \ 0.6$, (b) $\alpha = 0.999$ and $r=0.3$, (c) different set of $\alpha \  and \ r$. }
\label{fig: 4}
\end{figure}

\begin{figure}
\centering
  \includegraphics[width=8cm,height=5cm]{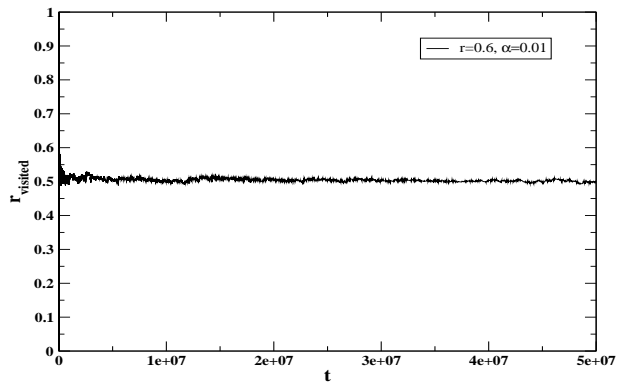}
  \caption{Plot of $r_{visited}$ vs time for $\alpha=0.01$ and $r=0.6$.}
  \label{fig: 6}
\end{figure}

When $\alpha $ is close to its boundary values (0 or 1), particle takes a much longer time to achieve an asymptotic behaviour than that of in the case of other values of $\alpha$. The asymptotic behaviour shows that particle performs diffusive motion for all values of $\alpha$ except when $\alpha=0$ and $\alpha =1.0$, in these cases particle motion is ballistic and confined respectively. When $\alpha \simeq 0.001$ and $r>0.5$, particle first move ballistically for early simulation time, then it shows much slower dynamics for significantly large time (say $\lambda$) and seems like the motion of particle is sub-diffusion, but, waiting further long, the dynamics again becomes faster and comparable to that in the case of normal diffusion; where MSD vs time exponent converges to $1$. It can be seen in fig. \ref{fig: 4}(a). Value of $\lambda$ increase as we decrease the value of $r$, i.e. the particle will take a longer time to show asymptotic behaviour for  $r<0.5$ than that of when $r>0.5$.  We explain this with the case when initially all the scatterers are of reflector type and no transmitter ($r=1.0$). In this case, we know if scatterers have pure flipping character $\delta=0$, then the particle will propagate with speed $v(t)=1/3$. Now if we have finite small $\delta$, scatterers have a small tendency to retain their character (non-flipping). In general, the particle is moving in the lattice with a speed of $1/3$. As soon as it encounters a non-flipping scatterer, it gets diverted from its original propagation direction and moves in the backward direction with speed $1/3$ to all the previously visited site.  Unless it again encounters a non-flipping scatterer and starts moving in the forward direction it started, with speed $1/3$.   This cycle keeps on and even the particle moves in propagating mode for some intermediate time, due to small non-flipping character whenever it encounters a non-flipping scatterer, it has to go a large distance in the backward direction, and hence the dynamics become slower. But it is never confined. Now if we tune  $\delta$ to moderate value, then it feels more random kicks from its velocity direction and motion tends to become more random and hence the MSD exponent $\beta$ increases and approaches to $1$ (diffusive type). During this process the particle also randomize the background lattice such that the value of $r$ on the visited site approaches to $0.5$ for $\alpha \neq 0 \neq 1$ i.e $r_{visited}=0.5$ as shown in fig. \ref{fig: 6}. Also, the rate of increase of the number of newly visited site $N(t)$ converges to zero therefore at long enough time, i.e. asymptotically system approaches to a stationary state where $r_{visited}=0.5$. It implies that at long enough time, the motion of the particle will be diffusive, i.e. the asymptotic behaviour is normal diffusion. We claim this behaviour as asymptotic since the rate of increase of newly visited site converges to zero, and the particle motion is randomising the background lattice during the total simulation time so that the system reaches to the stationary state. 

\begin{figure}
\centering
\includegraphics[width=15 cm,height=4.0 cm]{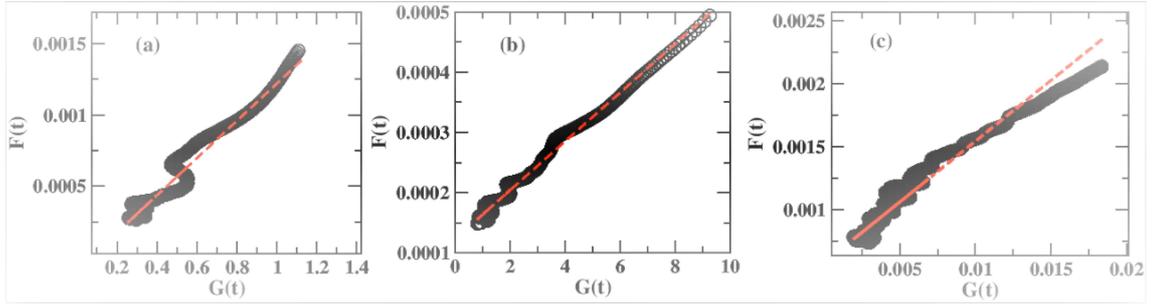}
\caption{$F(t)=\frac{d}{dt}N(t)$ vs. $G(t)=\frac{1}{N(t)}$ plot  when (a)$\alpha = 0.3$ and $r=0.3$ , (b) $\alpha = 0.003$ and $r=0.6$, (c)$\alpha = 0.003$ and $r=0.9$ . Data (black cilcles) is fitted linearly (red dashed line) which gives the slope $c=c(\alpha,r)$. }
\label{fig: 5}
\end{figure}

Asymptotic behaviour for $\alpha \simeq 0.999$ and $r<0.5$ can be understood in the similar way.
We have also seen that the number of different site visited $N(t)$ agrees well with equation (\ref{eq:1}) which have the solution $N(t)\propto \sqrt{t}$ which is the case of random walk in one dimension. Plots are given in fig. \ref{fig: 5} for different sets of parameters $(\alpha,r)$ where $F(t)=\frac{d}{dt}N(t)$ and $G(t)=\frac{1}{N(t)}$. Looking at these plots what we observe that the time derivative of number of different sites visited $\frac{d}{dt}N(t)$ and reciprocal of $N(t)$ are proportional to each other, hence, $N(t)$ approaches to zero at long time.

\begin{figure}
\centering
  \includegraphics[width=12cm,height=8cm]{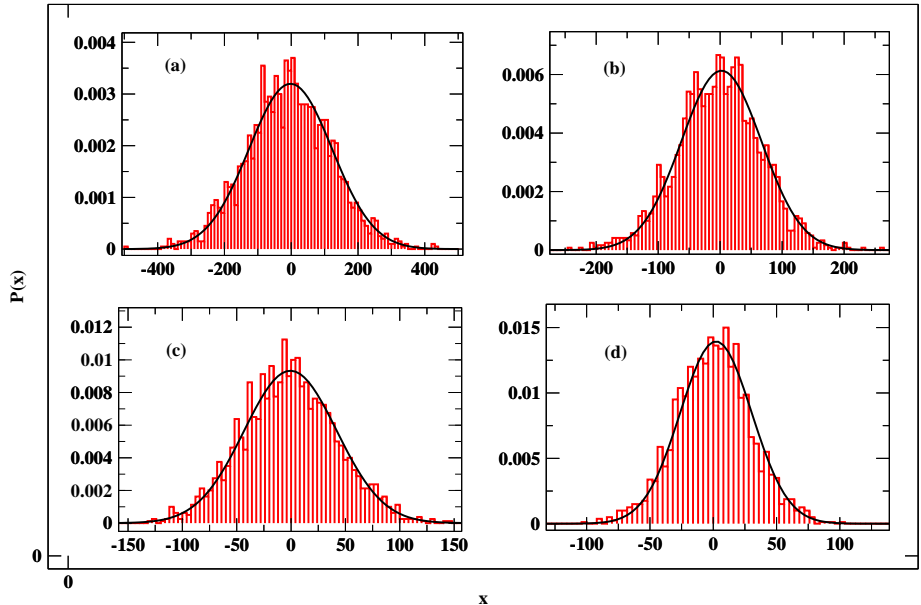}
\caption{Probability distribution of particles position (bars) and fitted with Gaussian distribution (solid line) when (a) $r=0.5$ and $\alpha =0.5$, (b) $r=0.5$ and $\alpha =0.9$, (c) $r=0.9$ and $\alpha =0.5$, (d) $r=0.9$ and $\alpha =0.9$.}
\label{fig: 7}
\end{figure}

{\em Anomalous diffusion}:-
We have calculated the probability distribution of particle's position for some chosen value of $\alpha$ and $r$ when the particle's dynamics is diffusive i.e. $\beta = 1.0$. Fig. \ref{fig: 7} shows the plot of probability distribution of particle position $P(x)$ , calculated for 2000 trajectories, for some choice of parameter set $(r, \alpha) = (0.5,0.5), (0.5,0.9), (0.9,0.5), (0.9,0.9)$. Bars are the data from the simulation and lines are fit to the Gaussian. Data fits well with the Gaussian. $P(x)$ is calculated by collecting particle's position at few random times for each $2000$ trajectories.\\
We also calculate the effective diffusion coefficient $D_{eff}$ using, $D_{eff}=\lim_{t \to \infty} \frac{\Delta(t)}{2t}$  in the numerical simulation and also estimate it as follows: consider the case, when $\alpha=0$ (pure flipping {\bf case 3}) typical path length of straight motion is $1/r$,  hence $1/r$ is like mean free path for pure flipping case. Now when we deviate $\alpha$ from $0$ ($\alpha > 0.0$), the average mean free path will decrease and typical speed of the particle when it moves in straight is $V \simeq (\frac{1-\alpha}{r}) \frac{1}{\tau}$ where  the rate $\tau$  at which particle changes its trajectory is $\frac{1}{\alpha}$. Hence estimated diffusion coefficient can be given by $D_{est} \simeq {V^2}{\tau} = \frac{(1-\alpha)^2}{r^2} \alpha$. The study of Gates {\it{et al.}} \cite{gates1982} also shows that diffusive behaviour of particle dynamics in the one-dimensional heterogeneous lattice. In our present study, we find change in the nature of particle dynamics, when we continuously change the underlying lattice properties, going from a completely deterministic to the probabilistic case.  In table \ref{table1}  we list the value of effective diffusivity $D_{eff}$ from simulation and estimated $D_{est.}$ for $\alpha=0.5$. Numerical data matches  well with estimated $D_{est.}$. The same argument do not hold as we go away from $\alpha=0.5$ as the randomness decreases by increasing $\alpha$. For $ r=0.5$, $D_{eff}$ approaches $0.5$, which is the value for one dimensional random walk. Now as we tune $r$, it can be tuned to larger values (for small r) and diverges for $r \rightarrow 0$ and to smaller values  (for large $r$) and approaches $1/8$ for $r=1$.

\begin{table}
\begin{center}
\caption{ List of the values of $D_{eff}$ and $D_{est}$ for different values of $r$ when $\alpha = 0.5$.}
 \label{table1}
  \begin{tabular}{l*{10}{c}r}
\hline
{\bf$r$}              & 0.1 & 0.2 & 0.3 & 0.4 & 0.5  & 0.6 & 0.7 & 0.8 & 0.9 & 1.0 \\
\hline
{\bf$D_{eff}$} & 12.142  & 3.434 & 1.592 & 0.770 & 0.467  & 0.335 & 0.268 & 0.212 & 0.172 & 0.123  \\
\hline
{\bf$D_{est}$}           & 12.500 & 3.125 & 1.389 & 0.781 & 0.500  & 0.347 & 0.255 & 0.195 & 0.154 & 0.125 \\
\hline
{\bf$\beta$}           & 01.003 & 1.067 & 1.087 & 1.003 & 1.022  & 1.029 & 1.060 & 0.960 & 0.984 & 1.004  \\
\hline
\end{tabular}
\end{center}
\end{table}

\begin{figure}
\centering 
\includegraphics[width=8cm,height=6cm]{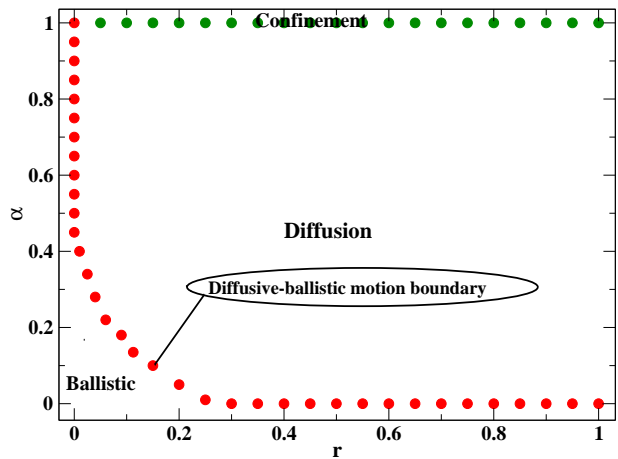}
\caption{Phase diagram: filled circles are data from simulation.}
  \label{fig: 8}
\end{figure}

{\em Phase Diagram}:-
We plot the phase diagram for the asymptotic behaviour of the particle motion, where the data obtained from the numerical simulation, in the plane of $\alpha - r$. Particle shows three different regimes of motion, i.e. ballistic, diffusion and confinement.  Phase boundaries between the three regimes of particle motion are drawn in the phase diagram shown in fig. \ref{fig: 8}.

\section{Discussion}
\label{discussion}
We have studied the dynamics of a particle moving in one-dimensional lattice-gas with randomly distributed reflectors and transmitters. The particle moves along the bonds of the lattice of unit spacing. Reflectors reflect the direction of particle velocity and transmitters leave it unchanged. Scatterers also change their character after interaction with the particle with probability $1-\alpha$. Hence for $\alpha=1$, nature of scatterers remain unchanged, and for $\alpha=0$, they always flip. Otherwise for $\alpha \ne 0 \ne 1$, flipping is probabilistic. Hence  $(r, \alpha)$ are the two control parameters in our model. \\
For $\alpha=0$ and $1$, dynamics of particle is purely deterministic: and it is completely confined or periodic for $\alpha=1$ and ballistic for $\alpha=0$ for all $r$. The region of confinement and speed of propagation depends on the initial value of $r$. For $\alpha \in (0,1)$, dynamics of particle shows a crossover from initial transient feature to late time steady state behaviour and asymptotically shows diffusive motion. Approach to the steady-state behaviour, in general, is quick and happens through a short transient state. But when the parameters $\alpha$ and $r$ are close to $0$ or $1.0$, particle takes a much longer time to reach its steady state and show asymptotic behaviour. In certain limit of the model, our results are in agrrement with previous study of \cite{kong1991, gates1982}. But our study is more general, here we give the full phase diagram in the plane of two relevant parameters. Hence our study motivate to extend such model in higher dimensions.

\section{Acknowledgement}
SK and SM would like to thank DST-INSPIRE Faculty award for financial support. SK would also like to thank  D. Giri,  Rajeev Singh, for their useful suggestions. SM would like to thank E. G. D. Cohen for introducing the problem of the dynamics of a particle on Lorentz lattice gas.

\end{document}